\documentclass[twoside,english,nofootinbib,preprintnumbers,aps,11pt,showpacs]{revtex4-1}
\usepackage{euscript,amssymb}
\usepackage{amsthm}
\usepackage{graphicx}
\usepackage{amsfonts}
\usepackage{amsmath}
\usepackage{amssymb}
\usepackage{fancyhdr}
\usepackage{epsfig}
\usepackage{color}
\usepackage{esint}
\usepackage{soul}
\usepackage{afterpage}
\usepackage[papersize={8.5in,11in}]{geometry}

\newcommand{\be}{\begin{equation}}
\newcommand{\ee}{\end{equation}}

\begin{document}

\title{Symmetries of the Trautman retarded radial coordinates}
\author{Maciej Kolanowski}
\email[]{mp.kolanowski@student.uw.edu.pl}
\author{Jerzy Lewandowski}
\email[]{Jerzy.Lewandowski@fuw.edu.pl}
 
\affiliation{\vspace{6pt} Institute of Theoretical Physics, Faculty of
  Physics, University of Warsaw, Pasteura 5, 02-093 Warsaw, Poland}
\begin{abstract}  We consider  spacetime described by observer that uses Trautman's retarded radial coordinates system. 
Given a metric tensor, we find all the symmetry maps. They set a 10 dimensional family of local diffeomorphisms 
of spacetime and  can be parametrized by the Poincar\'e algebra. This result is similar to the symmetries of a Gauss
observer using the Gauss simultaneous radial coordinates and experiencing the algebra  deformation induced by the spacetime 
Riemann tensor.   A  new, surprising property of the retarded coordinates is that while the symmetries  
are  differentiable, in general they are not differentiable twice. In other words, a family of smooth symmetries  is smaller than 
in the Gaussian case.  We demonstrate examples of that non-smoothness and find necessary conditions for the differentiability
to the second order.  We also discuss the consequences and relevance of that result for geometric relational observables program.  
One can interpret our result  as that the retarded radial coordinates provide gauge conditions stronger that the Gaussian  4d radial 
gauge.                     
\end{abstract}


\pacs{04.70.Bw, 04.50.Gh}

\maketitle

\section{Introduction}  In general relativity we often  use  coordinates  adapted in a natural way to 
a given metric tensor.  In that way we reduce the degrees of freedom in the metric tensor components 
related to  non-physical and non-geometrical ambiguities. Therefore one may thing of that choice of
coordinates as a gauge fixing. The diffeomorphisms  that destroy our gauge get eliminated. Usually,  though, 
there  still remains  a  family of diffeomorphisms that preserve the conditions imposed on  the coordinates.  
They are symmetries of a  framework based on that gauge choice.  An example of adapted coordinates may 
be Trautman - Bondi coordinates defined in asymptotically flat spacetime in a neighborhood of conformally defined 
boundary of spacetime \cite{Trautman_coordinates_infty,Bondi_coordinates,*Bondi_coordinates2}. Their  symmetries set  the famous  Bondi-Misner-Sachs (BMS) group. 
It is infinite dimensional, contains infinitely many subgroups isomorphic to the Poincar\'e group.  The relevance of  the 
BMS  group for generally relativistic physics is very well  known. In the current paper we consider coordinates adapted 
in a similar way to a timelike geodesic curve - an observer.  We use incoming null geodesic curves, proper time, affine 
parametrizations, and angles parametrizing the sphere of null directions parallely transported  along the observer's world line.  
Such coordinates were introduced for the first time  by  Trautman \cite{Trautman_coordinates}  in his search for special 
solutions to Einstein's equations, therefore we will name them after him. Our  work arises from the program of construction  
of  diffeomorphism invariant observables for  canonical general relativity naturally related to geometry
\cite{Duch_PhD}. This is a geometric version of the idea of the relational observables \cite{Relational_observables,*Relational_observables2,*Relational_observables3,*Relational_observables4}. 
In the previous papers  we considered  in that context  radial coordinates in 3d Euclidean space \cite{Radial_coordinates,*Radial_coordinates2,*Radial_coordinates3},  
and Gauss radial coordinates in 4d spacetime, respectively \cite{Gauss_coordinates}.  Similar construction was considered 
for  asymptotically AdS spacetimes \cite{ads1, *ads2, *ads3}. Analysis of (non-)locality of such observables was performed in perturbative approach\cite{Giddings}.  We found the symmetries of said coordinate systems and analyzed their structure \cite{Symmetries}.  
In the case of the radial coordinates in the  Euclidean  3d space, for every a metric tensor, the symmetries set a 6 dimensional 
family that may be parametrized by the Lie algebra of the isometries of Euclidean flat space.  In the case of the Gauss radial 
coordinates in 4d spacetime,  observer's  symmetries form a 10 dimensional family  parametrized by the Poincar\'e algebra. 
In the both cases, though,  the symmetries are not a group. Composition of two symmetries of a given metric 
tensor may happen not to be a symmetry any more. On the other hand, they do induce a group of motions of the space of the metric tensors.  
The generators of that group satisfy the commutation relations of the parameterizing Lie algebra suitably deformed by the 
Riemann curvature tensor (translations do not commute).  The Trautman coordinates  are more physical 
because of their retarded character. They are most difficult, though, from the point of view of the symplectic structure 
of Einstein's theory of gravity and other fields.  In the current paper we  investigate  symmetries of the Trautman 
coordinates.  A priori, one would expect that they again form a 10 dimensional family per a metric tensor parametrized
by the Poincar\'e algebra. We find that this guess is partially true with a catch, that the symmetries fail to be twice
differentiable. We provide examples, find some second differentiability conditions and discuss the relevance
of that surprising  result. 

\section{Trautman's retarded radial coordinates}
We consider a $4$-dimensional spacetime $M$ and metric tensor $g$ of the signature $-+++$.  Observer world line is a time-like curve 
in $M$
\be\label{obs}  \mathbb{R}\ni \tau \mapsto p(\tau)\in M \ee
where $\tau$ is the proper time, that is 
\be \frac{dp^\mu}{d\tau} \frac{dp_\mu}{d\tau}\ =\ -1.\ee
We are assuming throughout this paper that the observer world line  is geodesic
\be \nabla_{\frac{dp}{d\tau}} \frac{dp}{d\tau}\ =\ 0. \ee
Our considerations are local, for example we do not bring up the issue of extendability of the observer line. 

Trautman coordinates are defined in a neighborhood of the observer world line.  For each point $p(\tau)$ of the observer world line 
we consider the past null cone formed by incoming null geodesics (generators). On each null generator of the cone we introduce an affine 
coordinate $r$ growing to the past. Let $n^\mu$ be the corresponding null vector tangent to the cone and normalized such that
\be n^\mu\partial_\mu r = 1 . \ee
The ambiguity in the choice of  the $r$ coordinate is fixed by assuming the following two conditions:
\be r = 0 \ee 
at the  end of each generator (that is the intersection of the generator with the observer world line), and
\be n^\mu \frac{dp_\mu}{d\tau}\ =\ 1. \ee
Each null geodesic curve coming in a point at the observer world line can be characterized by the intersection point  
and two angle variables $\theta^A = \theta,\phi$ parameterizing the sphere of null directions at that point. We assume that 
the angles $\theta^A$ are constant with respect to the parallel transport of the null vectors along
the observer world line.   Now, to every point $m$ in a sufficiently narrow neighborhood of the observer world line 
(and not contained in the world line itself) there corresponds uniquely: a null curve connecting $m$ with the observer 
world line, the value $\tau(m)$ 
of the observer time at the intersection point, the value $r(m)$ of the affine parameter corresponding
to the point $m$, and the angles $\theta(m), \phi(m)$ characterizing the null geodesic curve. In that way we 
have introduced a coordinate system $(\tau,r,\theta,\phi)$, 
\be m\ \mapsto\ (\tau(m),r(m),\theta(m),\phi(m)) \ee
in the neighborhood of the observer world line that is singular at the world line. We call that neighborhood 
the domain of the coordinates (despite the singularity at $r=0$).  In those 
coordinates the metric tensor components satisfy a very simple (gauge) condition
\be \label{gauge} g_{r\mu}\ =\ \delta_{\tau\mu}. \ee
We have to remember, however, about additional very non-trivial (gauge) conditions implied by the above construction 
and by the differentiability of the metric tensor $g$ - for simplicity let us assume $g$  is smooth, but in fact we will use only the second derivatives 
of $g$ to calculate it's  Riemann tensor - on the components $g_{\tau\tau},g_{\tau A}, g_{AB}$  and their derivatives 
to arbitrary order  in the limit  as  $r\mapsto 0$.   

Given a metric tensor $g$, each system of Trautman coordinates  is determined by
a point $m_0\in M$ and an orthonormal tangent frame $e_0,e_1,e_2,e_3\in T_{m_0}M$. Indeed, 
the observer world line is determined by the time like vector $e_0$, the proper time $\tau$ is chosen to vanish  at $m_0$, 
the vectors $e_1,e_2,e_3$  determine in a standard way a  spherical coordinates system $\theta,\phi$ on the 
sphere of the null directions at $m_0$. 
This correspondence
\be  (\tau,r,\theta,\phi)\ \leftrightarrow (m_0,e_0,e_1,e_2,e_3) \ee
is 1-to-1. 
\section{Symmetries} Given a metric tensor $g$ and  Trautman coordinates 
$(\tau,r,\theta,\phi)$, a symmetry is a local diffeomorphism $f$ defined in an open and connected region contained in  the 
domain of the coordinates and containing a segment of the observer line, such that  for the pulled back  metric tensor $f^*g$ 
the functions $(\tau,r,\theta,\phi)$ also satisfy the definition of Trautman coordinates.   In other words, $f$ is a symmetry 
whenever the metric tensor $f^*g$ also satisfies the gauge conditions (\ref{gauge}) and the conditions described below  (\ref{gauge}).    

We  find now all the symmetries modulo a catch we will point out at the end. 

Suppose $f$ is a symmetry. Then, the functions ${f^{-1}}^*\tau,{f^{-1}}^*r,{f^{-1}}^*\theta, {f^{-1}}^*\phi$ set another system of retarded radial coordinates
for the metric tensor $g$.  Therefore, every symmetry of a given Trautman coordinates  $(\tau,r,\theta,\phi)$ can be obtained in the following way:
let $(\tau',r',\theta',\phi')$ be another system of Trautman coordinates for the metric
tensor $g$. Define $f$ by the equation 
\be\label{symmap} (\tau(m), r(m), \theta(m),\phi(m))\ =\ (\tau'(f(m)), r'(f(m)), \theta'(f(m)),\phi'(f(m))) .\ee
Conversely, every map $f$ defined in that way  is a symmetry of $g$ and $(\tau,r,\theta,\phi)$. 
 
Now, the catch is the differentiability level of the map $f$ at the points of the observer world line. Indeed,
at the points of $M$ such that $r>0$  ($r'>0$), the coordinates  $(\tau,r,\theta,\phi)$   ($(\tau',r',\theta',\phi')$), 
and in the consequence $f$, are all smooth. At the world line, on the other hand, our Trautman coordinates are not very useful
to judge about the differentiability. However,  at a point $p(\tau_1)$ on the observer world line (\ref{obs}), every vector 
$X\in T_{p(\tau_1)}M$ can be uniquely written as
\be  X \ =\ a \frac{dp}{d\tau}(\tau_1) + b \partial_r^{(\theta_1,\phi_1)}, \ \ \ \ \ \ \ a,b\in\mathbb{R}\ee
where $\partial_r^{(\theta_1,\phi_1)}$ is the limit $r\mapsto 0$ of the vector field $\partial_r$ along a 
null geodesic curve corresponding to  $(\tau_1,\theta_1,\phi_1)$. Every curve tangent to $X$ is mapped
by the symmetry $f$ to a curve tangent to  a vector   
\be X'\ =\ a \frac{df(p)}{d\tau'}(\tau_1) + b \partial_{r'}^{(\theta_1,\phi_1)}.\ee  
It is easy to see that this map 
\be T_{p(\tau_1)}M \ni  a \frac{dp}{d\tau}(\tau_1) + b \partial_r^{(\theta_1,\phi_1)}\ \mapsto\ 
a \frac{df(p)}{d\tau'}(\tau_1) + b \partial_{r'}^{(\theta_1,\phi_1)}\in   T_{f(p)(\tau_1)}M  \ee
is linear and actually it is the derivative $f_*$ of the map $f$ at $p(\tau_1)$.  

In summary, given a metric tensor $g$ and Trautman coordinates $(\tau,r,\theta,\phi)$, we have defined
and characterized the  10 dimensional  family  of the symmetries. Obviously, 
the symmetries depend on the coordinates $(\tau,r,\theta,\phi)$.   What they essentially depend on
is the underlying  observer world line $p$, modulo the beginning point $p(0)$.  Therefore we can denote them
${\rm Sym}(g,p)$.  Notice, that if $g\not=g'$ then in general   
\be {\rm Sym}(g',p) \not= {\rm Sym}(g,p). \ee
For that reason, ${\rm Sym}(g,p)$ is not a group except when $g$ is Minkowski, de Sitter, or anti-de Sitter.   
The symmetries  are everywhere differentiable at least once. In general though,  a symmetry of Trautman 
coordinates  is not twice differentiable at the observer line  (see below).       

\section{Example of a non-C$^2$ symmetry}  
In this section we give an example of a metric tensor $\tilde{g}$, Trautman coordinates $(\tilde{\tau},\tilde{r},\tilde{\theta},\tilde{\phi})$ 
and a symmetry $\tilde{f}$ defined by another Trautman coordinates system  $(\tilde{\tau}',\tilde{r}',\tilde{\theta}',\tilde{\phi}')$ 
that is not twice differentiable at points on  the observer world line. 

The construction of our example is suggested by the  following observation. 
Given a metric tensor $g$ and Trautman coordinates $(\tau,r,\theta,\phi)$, one may introduce  Cartesian  coordinates 
$(T,X,Y,Z)=(T, X^i)$ defined as follows \begin{align} 
       T =& \tau - r\nonumber\\
       X =& r{\rm sin}\theta{\rm cos}\phi \nonumber\\
       Y =&  r{\rm sin}\theta{\rm sin}\phi\nonumber\\
       Z=&  r{\rm cos}\theta
       \label{Cart}
 \end{align}
In terms of them, the  formula (\ref{symmap}) for  a symmetry $f$ takes an analogous form
\be\label{symmapcar} (T(m), X(m), Y(m),Z(m))\ =\ (T'(f(m)), X'(f(m)), Y'(f(m)),Z'(f(m))) \ee
where $(T',X',Y',Z')$ are the Cartesian coordinates corresponding to the primed Trautman coordinates.  
Now, suppose, that the metric tensor $g$ is flat in a  neighborhood of the primed observer's world line, 
contained in the domain of the primed coordinates $(\tau',r',\theta',\phi')$. Then, the primed  Cartesian 
coordinates $(T',X',Y',Z')$ are smooth  including the  primed world line $r'=0$. Obviously, in terms of them 
the metric tensor is just   
\be\label{flat}  g\ =\ -dT'^2 + dX'^2 + dY'^2 + dZ'^2\ee
in the flatness region. It follows from (\ref{symmapcar}), that at every point
$m$, the symmetry $f$ is exactly as many times differentiable as the map
\be m\mapsto (T(m), X(m), Y(m),Z(m)) . \ee 
Therefore, to construct an example of at most once differentiable symmetry it is
sufficient to find a metric tensor $g$ and retarded radial coordinates  $(\tau,r,\theta,\phi)$ such that 
the corresponding Cartesian functions are at most once differentiable. Then, the example will be given
by gluing a neighborhood of  observer's world line of  the Trautman coordinates    
$(\tau,r,\theta,\phi)$  endowed with the metric $g$, with a neighborhood of observer's world line
of the Trautman coordinates  $(\tau',r',\theta',\phi')$ endowed with the flat metric $g'$.  
The gluing is such that the neighborhoods become  disjoint  neighborhoods in a single
spacetime $M$, and a metric $\tilde{g}$ is such that in one neighborhood $\tilde{g}=g$
and  in the other one $g=g'$. The corresponding symmetry $\tilde{f}=f$  given by (\ref{symmap}).  

We  give now an example of a metric $g$ and Trautman coordinates $(\tau,r,\theta,\phi)$
such that the corresponding Cartesian coordinates are not differentiable more than once. 
\newline
Let us consider FLRW spacetime filled with radiation. In the standard cartesian coordinates $(t, x, y,z)$ the metric 
takes following form
\be  g = -dt^2 + t \left(dx^2 + dy^2 + dz^2 \right)\ee
One can easily integrate the null geodesic equation. Without loss of generality consider it in  the $y=0=z$ plane. 
We have following equations
\begin{equation}
\begin{cases}
\left(\frac{dt}{dr}\right)^2 = t\left(\frac{dx}{dr}\right)^2 \\
t\frac{dr}{d\lambda} = const. = 
\sqrt{
\tau} 
\end{cases}
\end{equation}
where $r$ is an affine parameter, the second line comes from properties of Killing vector $\partial_x$ and $\tau$ is the time when geodesic crosses observer's word line. By direct integration we obtain
\begin{equation}
\begin{cases}
t = \tau \left(1 - \frac{3r}{2\tau}\right)^{\frac{2}{3}} \\
x = -2\sqrt{\tau} \left(1 - \frac{3}{2}\frac{r}{\tau}     \right)^{\frac{1}{3}}  + 2\sqrt{\tau}
\end{cases}
\label{geo}
\end{equation}
Generalization for any radial coordinate is straightforward. Non-zero elements of metric in those new coordinates $(\tau, r, \theta, \phi)$ are as follows
\begin{equation}
\begin{cases}
g_{\tau \tau} = \frac{r}{\tau}-\left(1-\frac{3 r}{2 \tau}\right)^{2/3}-2 \\
g_{\tau r} = g_{r \tau} = 1 \\
g_{\theta \theta} = \tau^2 \left(1-\frac{3 r }{2 \tau}\right)^{2/3} \left(2 \sqrt[3]{1-\frac{3 r }{2 \tau}}-2\right)^2 \\
g_{\phi \phi} = \tau^2 \left(1-\frac{3r }{2\tau}\right)^{2/3} \left(2 \sqrt[3]{1-\frac{3 r }{2 \tau}}-2\right)^2 \sin ^2(\theta)
\end{cases}
\end{equation}
It is easy to see that the gauge conditions (\ref{gauge}) are satisfied.
We can also use obtained results to find associated Cartesian coordinates. Firstly, one can invert \eqref{geo}
\be \begin{cases} \tau= \frac{(x+2\sqrt{t})^2}{4}      \\
 r = \frac{2}{3} \left(\frac{(x+2\sqrt{t})^2}{4} - \frac{2t\sqrt{t}}{x+2\sqrt{t}}     \right)  \end{cases}    \ee
which, using \eqref{Cart}, gives us
\begin{equation}
\begin{cases}
T = \tau - r = \frac{(\sqrt{x^2+y^2+z^2}+2\sqrt{t})^2}{12} + \frac{4}{3} \frac{t\sqrt{t}}{\sqrt{x^2+y^2+z^2}+2\sqrt{t}} \\
X^i = r \frac{x^i}{x^2 +y^2 + z^2} = \frac{1}{6}x^i \frac{x^2+y^2+z^2 + 6\sqrt{x^2+y^2+z^2}\sqrt{t}+12t}{\sqrt{x^2+y^2+z^2} +2\sqrt{t}} = \sqrt{t}x^i + x^i \frac{x^2+y^2+z^2}{6\sqrt{x^2+y^2+z^2} +12\sqrt{t}}
\end{cases}
\end{equation}
The non--smoothness at $x=y=z=0$ follows from the presence of square root in denominator. Specifically, each of the functions 
$T,X,Y,Z$ is everywhere differentiable in accordance with the conclusion of the previous section.  
But consider the  mixed second derivative:
\be \begin{split}
\frac{\partial^2 T}{\partial x \partial y} = \frac{xy\sqrt{t}}{3(x^2+y^2+z^2)^{\frac{3}{2}}}  \left(\frac{4t}{(2\sqrt{t}+(x^2+y^2+z^2)^{\frac{1}{2}})^2} -1\right) + \\
\frac{8 t \sqrt{t}}{3(2\sqrt{t}+\sqrt{x^2+y^2+z^2})^2} \frac{xy}{\sqrt{x^2+y^2+z^2}} \left(1 + \frac{2 \sqrt{t}}{\sqrt{x^2+y^2+z^2}} \right)
\end{split}
\ee
The very last term in this whole expansion is the only one which is of order $t$ (the rest is of order $\sqrt{t}$) and it is clearly non-continuous so our coordinates are not $C^2$.
\section{The second differentiability conditions in  the flat observer case}
In this section we continue the study  of the  symmetries and their construction via the formulae   (\ref{symmap})
and (\ref{symmapcar}).  As in the example of the previous section, we again assume that the metric $g$ is flat in a neighborhood of the 
primed observer world line where the primed Cartesian coordinates are used, that is (\ref{flat}) holds there again.  However, now we find  
necessary conditions on $g$ (and its Riemann tensor) that hold at the unprimed observer world line for  the existence of the second derivative 
of the corresponding symmetry $f$   at observer's world line. The condition we derive below reads
\be\label{thecond} R_{\alpha\beta\gamma\delta}\frac{dp^\delta}{d\tau}|_{p(\tau)}\ =\ 0 ,  \ee
for every point $p(\tau)$ of the world line, where $R^\alpha{}{}_{\beta\gamma\delta}$ is the Riemann tensor of $g$.  
We show below the calculation that leads to that conclusion.

 If the cartesian coordinates $(T,X,Y,Z)$ (\ref{Cart}) are $C^2$, then so are  $g_{IJ}$ as functions of these coordinates.  In terms of those
 coordinates the  gauge conditions  (\ref{gauge}) read as follows
\begin{equation} \begin{cases}
-g_{TT} + g_{TX^i} \frac{X^i}{r} = -1 \\
g_{TT} + g_{ij} \frac{X^i X^j}{r^2} -2g_{TX^i} \frac{X^i}{r} = 0 \\ 
\frac{Z}{\rho} \left(-g_{TX}X - g_{TY}Y \right) + g_{TZ}\rho + \frac{Z}{\rho r} \left(g_{XX} X^2 +2g_{XY}XY +g_{YY}Y^2 \right) + \\ \frac{-X^2-Y^2+Z^2}{\rho r} \left(g_{XZ}X + g_{YZ} Y \right) - g_{ZZ} \frac{Z \rho}{r} = 0 \\
g_{TX} Y - g_{TY}X + \left(g_{YY}- g_{XX}\right)\frac{XY}{r}+g_{XY}\frac{X^2-Y^2}{r} - g_{XZ}\frac{YZ}{r} + g_{YZ}\frac{XZ}{r} = 0
\end{cases} \label{gauge2} \end{equation}
where $r = \sqrt{X^2 + Y^2 + Z^2}$ and $\rho = \sqrt{X^2 + Y^2}$.
Consider the Taylor expansion of the metric around the unprimed observer world line ($r=0$), namely
\[ g_{IJ} = \eta_{IJ} + k_{IJ} + h_{IJ} + O(r^3)          \]
where $k$ and $h$ are homogeneous polynomials in $X,Y,Z$ of the order 1 and 2, respectively. Their coefficients can depend on T. We can apply this expansion to \eqref{gauge2} and separate equations according to the orders in question. Also, one can easily notice that $k$ and $h$ are odd and even, respectively, under the parity transformation. This allows us to separate those conditions even more. As a result, one obtains the following linear system in which the $k_{IJ}$ and $h_{IJ}$ are
considered as unknown, and $X,Y,Z,\rho$ as given
\begin{equation}
\begin{cases}
k_{TT} = h_{TT} = 0 \\
k_{TX^i} X^i = 0 = h_{TX^i} X^i \\
-k_{TX}XZ - k_{TY}YZ + k_{TZ} \rho^2 = 0 = -h_{TX}XZ - h_{TY}YZ + h_{TZ} \rho^2 \\
k_{TX}Y - k_{TY}X = 0=h_{TX}Y - h_{TY}X
\end{cases}
\label{ukl1}
\end{equation}
\begin{equation}
\begin{cases}
k_{ij}X^i X^j = 0 = h_{ij}X^i X^j \\
k_{XX} X^2 Z + 2 k_{XY} XYZ + k_{YY}Y^2Z - k_{ZZ} Z \rho^2 - (X^2+Y^2-Z^2) \left(k_{XZ} X + k_{YZ} Y \right) =0 \\
h_{XX} X^2 Z + 2 h_{XY} XYZ + h_{YY}Y^2Z - h_{ZZ} Z \rho^2 - (X^2+Y^2-Z^2) \left(h_{XZ} X + h_{YZ} Y \right) =0 \\
\left(k_{YY}-k_{XX}\right)XY + k_{XY}\left(X^2-Y^2\right) - k_{XZ}ZY + k_{YZ} XZ = 0 \\
\left(h_{YY}-h_{XX}\right)XY + h_{XY}\left(X^2-Y^2\right) - h_{XZ}ZY + h_{YZ} XZ = 0
\label{ukl2}
\end{cases}
\end{equation}
\eqref{ukl1} is a Cramer, homogeneous system of equations so it has only trivial solution. However, \eqref{ukl2} admits non-trivial following solution
\begin{equation}\begin{cases}
h_{XX} =  aY^2 + 2b Y Z + c Z^2 \\
h_{XY} = -aXY - bXZ + dYZ + e Z^2 \\
h_{XZ} = - bXY - c XZ - dY^2 - e YZ \\
h_{YY} = a X^2 - 2dXZ + fZ^2 \\
h_{YZ} = bX^2 + dXY - eXZ - fYZ \\
h_{ZZ} = cX^2 + 2eXY + fY^2
\end{cases}
\end{equation}
where $a,b,c,d,e,f$ are arbitrary T-dependent functions.  The Riemann tensor of this metric necessarily satisfies at $r=0$
\be R_{T\beta \gamma \delta}{}_{|_{r=0}} = R_{\alpha \beta \gamma \delta} \frac{dp^\alpha}{d\tau} = 0 \ee 
as it is stated above. 

One can also find the non-vanishing components of Riemann tensor at $r=0$. Those are
\begin{align} R_{XYXY} &= \frac{1}{2} \left(-a - a - 2a -2a \right) = -3a  \nonumber    \\
 R_{XYXZ} &= \frac{1}{2} \left(-b-b-2b-2b\right) = -3b   \nonumber  \\
 R_{XYYZ} &= \frac{1}{2} \left(-2d-2d-d-d \right) = -3d    \nonumber         \\
 R_{XZXZ} &= \frac{1}{2} \left(-c - c - 2c-2c \right) = -3c     \nonumber \\
 R_{XZYZ} &=\frac{1}{2} \left(-e-e-2e-2e\right) = -3e \nonumber   \\
 R_{YZYZ}  &= \frac{1}{2} \left(-f-f-2f-2f \right) = -3f    \end{align}
There are exactly six of them which is the number of independent components of a 3-dimensional Riemann tensor. So in general,  $R_{\alpha \beta \gamma \delta}$ pulled-back to the surface $T=const.$ can be arbitrary. These conditions are satisfied in particular by any ultrastatic spacetime. \\
From physical point of view, the object of interest is Einstein tensor which takes following form
\begin{equation} G_{IJ} = R_{IJ} - \frac{1}{2}Rg_{IJ} =-3 \left(\begin{array}{cccc} 
a+c+f & 0 & 0 & 0\\ 
0 & -f & e & -d \\ 
0 & e & -c & b \\ 
0 & -d & b & -a   
\end{array} \right)    \end{equation}
Since all diagonal terms cannot be positive at the same time, matter coupled to gravity in this case needs to be exotic one. Vacuum, on the other hand, implies vanishing of Riemann tensor at the world line of the observer.


\section{Discussion}  Our result can be discussed at several levels.  We start from a direct technical interpretation,  and proceed toward
more specialized ones. 

Technically, what we have discovered leads to a relation between spacetimes endowed with the Trautman coordinates:
$(g,\tau,r,\theta,\phi)$ and $(g',\tau', r', \theta',\phi' )$ are in the relation provided the map defined by the  coordinates via (\ref{symmap}) is smooth. We have showed that the map is always differentiable. However, we gave an example of a pair such
that the map is not differentiable for the second time.  The example was followed by a more general condition for the pair to be in the 
relation if $g'$ is flat, namely then,  a necessary condition is a constraint on the Riemann tensor at the observer's line (\ref{obs})
$$ \frac{dp^{\mu_1}(\tau)}{d\tau}R_{\mu_1 \mu_2\mu_3\mu_4}{}_{|_{r=0}} = 0, $$
where the indices $\mu_1,...,\mu_4$ refer to any coordinate system non-singular   at $r=0$. 
A general necessary and sufficient condition is not known to us.    

From the gauge fixing point of view, the symmetries of adapted coordinates are the residual diffeomorphisms that  
are not eliminated by our gauge choice.   Hence, they are a failler of the corresponding gauge. In the current case 
of the Trautman coordinates the family of symmetries is for every metric tensor $10$ dimensional, 
however in a general case a symmetry is differentiable only once, and not necessarily twice.    Therefore, if one insists
on the smoothness of the symmetries, the conclusion is that   their family is smaller than $10$-dimensional. Hence, the  
lack of the smoothness we have encountered makes the  gauge choice corresponding to the 
Trautman coordinates  actually better, than say the Gauss radial coordinates. An exact dimension of the family of
the symmetries is not known at this point. For that we would need a general sufficient and necessary condition for the 
smootheness of the map (\ref{symmap}).    

Finally, we can look at the the issue from the point of view of the observer that  describes the surrounding spacetime
by using the Trautman coordinates. The observer has at her or his disposal the set $M_{\rm obs}$ of values the quadruples $(\tau,r,\theta,\phi)$ 
of the coordinates take. The spacetime geometries emerge to the observer in the form of  the family of  metric tensors that 
satisfy (\ref{gauge}) and the complicated conditions at $r=0$. In $M_{\rm obs}$ the smooth  manifold structure breaks at the points 
corresponding to $r=0$. The observer may gain insight into that structure by introducing the Cartesian coordinates system 
(\ref{Cart}) corresponding to some metric tensor $g$.   According to our results, those coordinates recover correctly the topological 
and differential (to the first order) structure of spacetime  at $r=0$. However, the metric tensor components in those coordinates 
will in general fail to be twice  differentiable. That is indication that the Cartesian coordinates do not encode  the spacetime smooth 
structure.   The observer may  analyze a structure 
of the singularity of the second and higher derivatives of $g_{IJ}$ and turn it into a set of new laws of physics. In particular she or he will notice, 
 that whenever the coordinates are one more time differentiable at $r=0$, then the 
Riemann tensor satisfies (\ref{thecond}).  In a matter of fact,  if $g$ is flat in a neighborhood of the line $r=0$, than 
the smooth structure defined in $M_{\rm obs}$ by the Cartesian coordinates agrees with that of an actual spacetime.  
 On the other hand, for arbitrary metric tensor there always exists some improved  coordinates 
 $\tilde{X}^\mu =\tilde{T},\tilde{X},\tilde{Y},\tilde{Z}$ dependent on the metric tensor $g$, non-smooth at $r=0$, making  the metric 
 tensor
  $$ g_{\mu\nu}(\tilde{T},\tilde{X},\tilde{Y},\tilde{Z})d\tilde{X}^\mu d\tilde{X}^\nu$$
smooth  at $r=0$.  Those coordinates reflect the actual manifold structure including the points 
of $r=0$.  That structure in observer's space $M_{\rm obs}$, however, depends on the metric tensor. 

\section*{Acknowledgements}

This work was partially supported by the Polish National Science Centre grant \newline No.~2011/02/A/ST2/00300.

\providecommand{\newblock}{}

\end{document}